\documentclass[showkeys,superscriptaddress,floatfix,prd,10pt,aps]{revtex4-2}
\usepackage{graphicx,epstopdf}
\usepackage[caption=false]{subfig}
\usepackage{appendix}
\usepackage[T1]{fontenc}
\usepackage{lmodern}
\usepackage[dvipsnames,x11names]{xcolor}
\usepackage[colorlinks=true,linkcolor=NavyBlue,citecolor=ForestGreen,urlcolor=NavyBlue]{hyperref}
\usepackage[sort&compress]{natbib}
\usepackage{lipsum}
\usepackage{morefloats}
\usepackage[pdf]{pstricks}
\usepackage{amsmath}
\usepackage{amssymb}
\usepackage{amsfonts}
\usepackage{rotating}
\usepackage{cancel}
\usepackage{mathtools}
\usepackage{bbm}
\usepackage{dsfont}
\usepackage{bbold}
\usepackage{multirow}
\usepackage{ulem}
\usepackage{physics}
\usepackage{orcidlink}
\usepackage{colortbl}

\def\pslash{p\!\!\!\slash }
\def\p_1slash{p1\!\!\!\slash }
\def\p_2slash{p2\!\!\!\slash }
\def\qslash{q\!\!\!\slash }
\def\xslash{x\!\!\!\slash }
\def\eslash{\varepsilon\!\!\!\slash }

\def\vel{\left|}
\def\ver{\right|}

\begin{document}

\title{Elucidating the nature of hidden-charm pentaquark states with spin-$\frac{3}{2}$ through their electromagnetic form factors}

\author{Ula\c{s}~\"{O}zdem\orcidlink{0000-0002-1907-2894}}%
\email[]{ulasozdem@aydin.edu.tr }
\affiliation{ Health Services Vocational School of Higher Education, Istanbul Aydin University, Sefakoy-Kucukcekmece, 34295 Istanbul, T\"{u}rkiye}

\date{\today}

\begin{abstract}
We perform a systematic study of the electromagnetic properties of exotic states to shed light on their nature, which is still controversial and not fully understood.  The magnetic dipole and higher multipole moments of a hadronic state are as fundamental a dynamical quantity as its mass, and they contain valuable information about the deep structure underlying it.  In the present work, we have explored the magnetic dipole and higher multipole moments of the hidden-charm pentaquarks with quantum number $J^P = 3/2^-$ using the QCD light-cone sum rule method and different interpolating currents. The obtained results show that different interpolating currents employed to probe pentaquarks with the same quark content produce varying results for their magnetic dipole and higher multipole moments at all.   This can be interpreted to mean that there is more than one hidden-charm pentaquark with identical quark content but with different magnetic dipole and higher multipole moments.  The nature, internal structure, and quark-gluon configurations of these states can be better understood by studying their electromagnetic properties.
\end{abstract}

\maketitle

\section{motivation}\label{motivation}

 The recent developments in experimental research have led to numerous new observations of both conventional and nonconventional hadrons. The adventure with unconventional hadrons began in 2003 with the discovery of the X(3872) state by the Belle Collaboration and reached another breaking point in 2015 with the discovery of hidden-charm pentaquarks by the LHCb Collaboration. Since then, a large number of unconventional hadron states have been observed experimentally and, together with theoretical studies, remain the most active and interesting area of particle physics.  Improved experimental techniques and analyses have led to the observation of unconventional states with increasing confidence levels. These investigations have enhanced our understanding of the strong interaction and the properties of these observed states.

In 2015, the LHCb collaboration first observed the pentaquark state with hidden-charm, which appeared to be inconsistent with the predictions of the conventional quark model. Since then, experimental collaborations have shown great interest in the pentaquark states with hidden-charm. An increasing amount of data has produced many hidden-charm pentaquark states. So far, the hidden-charm pentaquarks $P_{c }(4380)$, $P_{c }(4450)$, $P_{c}(4312)$, $P_{c}(4440)$, $P_{c}(4457)$, $P_c(4337)$, $P_{cs}(4459)$, and $P_{cs}(4338)$ have been experimentally observed~\cite{Aaij:2015tga, Aaij:2019vzc, Aaij:2020gdg, LHCb:2021chn, LHCb:2022ogu}. The spin-parity quantum numbers of the pentaquark state with hidden-charm states have been predicted by many theorists since their discovery in experiments. However, reliable information about their spin-parity has been unavailable until now~\cite{Esposito:2014rxa,Esposito:2016noz,Olsen:2017bmm,Lebed:2016hpi,Nielsen:2009uh,Brambilla:2019esw,Agaev:2020zad,Chen:2016qju,Ali:2017jda,Guo:2017jvc,Liu:2019zoy,Yang:2020atz,Dong:2021juy,Dong:2021bvy,Chen:2022asf,Meng:2022ozq}.  One of the remaining challenges to be solved is to understand the inner structure of the experimentally discovered pentaquark states with the hidden charm.

As previously stated, numerous investigations have been conducted to clarify the inner structure and exact nature of these nonconventional states. However, their inner structures remain incompletely understood.  Studying various properties together with masses and decay channels can provide insight into their characteristics and internal structure.  
The electromagnetic properties of hadrons play a significant role in their strong interactions and structure, providing insight into QCD in the low-energy regime. These observables can provide valuable information about the distributions of charge and magnetization, as well as their geometric shape. The literature contains few studies investigating the electromagnetic form factors of pentaquark states with hidden-charm, despite the valuable information they provide. The electromagnetic features of the pentaquark state with hidden-charm were theoretically investigated in various configurations and models~\cite{Wang:2016dzu, Ozdem:2018qeh, Ortiz-Pacheco:2018ccl,Xu:2020flp,Ozdem:2021btf,Ozdem:2021ugy, Li:2021ryu,Ozdem:2023htj,Wang:2023iox,Ozdem:2022kei,Guo:2023fih,Ozdem:2022iqk,Wang:2022nqs,Wang:2022tib,Ozdem:2024jty,Li:2024wxr,Li:2024jlq}. In the future, the discovery of many new pentaquarks is not far off with the continuous advancement of experimental techniques. Researchers will be more interested in exploring information about the internal structure of the pentaquark states, and exploring the magnetic dipole and higher multipole moments can provide a significant amount of reference data.  Motivated by this, in this work, in the compact diquark-diquark-antiquark pentaquark picture, the magnetic dipole and higher multipole moments of the pentaquark states with hidden-charm are studied using the QCD light-cone sum rule technique.

Before starting the analysis, it is important to briefly discuss how to measure the magnetic dipole moment of the hidden-charm pentaquark states in the experiment. Measuring the magnetic dipole moments of hidden-charm pentaquarks through spin precession experiments is challenging due to their short lifetimes.  Instead, the magnetic dipole moment of this unstable state can only be measured indirectly through a three-step process. First, the particle is produced. Then, it emits a low-energy photon which acts as an external magnetic field. Finally, the particle decays.   The pentaquark states with hidden-charm can be produced through either the photo-production or the electro-production process $\gamma^{(*)}N$ $\rightarrow$ $P_{c}$ $\rightarrow$ $ J /\psi N$.  Another challenging process $\gamma^{(*)}N $ $ \rightarrow $ $P_{c} $ $\rightarrow P_{c} \gamma$ $ \rightarrow $ $ J /\psi N \gamma$ can be analyzed to extract the magnetic dipole moments of the pentaquark states with hidden-charm.   The magnetic dipole moment of $\Delta$ resonance was acquired from a very similar $\gamma N $ $ \rightarrow $ $ \Delta $ $\rightarrow $ $ \Delta \gamma $ $ \rightarrow$ $ \pi N \gamma $ process~\cite{Pascalutsa:2004je, Pascalutsa:2005vq, Pascalutsa:2007wb,   Kotulla:2002cg,Drechsel:2001qu,Machavariani:1999fr,Drechsel:2000um,Chiang:2004pw,Machavariani:2005vn}.   
The cross sections, whether total or differential, may be influenced by the magnetic dipole moment of pentaquark states. The extraction magnetic dipole moment of these states can be achieved by comparing theoretical predictions with measured cross-sections. However, it should be noted that extracting the magnetic dipole moments of the hidden-charm pentaquark states using this technique presents a significant challenge.  

The work is outlined as follows: The next section provides details on the QCD light-cone sum rule calculations for the magnetic dipole and higher multipole moments of all considered states. Section~\ref{numerical} presents the analysis and discussion of the numerical results.  The summary and conclusion can be found in Section~
\ref{summary}.

\begin{widetext}
 
\section{QCD light-cone sum rule for magnetic dipole and higher multipole moments}\label{formalism}

The correlation function needed to calculate the magnetic dipole and higher multipole moments is given by the following equation
\begin{eqnarray} \label{edmn01}
\Pi_{\mu \nu}(p,q)&=&i\int d^4x e^{ip \cdot x} \langle0|T\left\{J_{\mu}(x)\bar{J}_{\nu}(0)\right\}|0\rangle _\gamma \, ,
\end{eqnarray}
where  the $J_{\mu}(x)$ stands for interpolating currents of the hidden-charm pentaquark states and  $\gamma$ is the external electromagnetic field. 

To understand the structure and properties of a hadron, one effective approach is to assign a suitable structure and calculate its properties. Theoretical findings resulting from these calculations can significantly aid in understanding the nature and substructure of the hadron.  To complete this task, it is crucial to choose appropriate interpolating currents that consist of quark fields aligned with the valence quark content and quantum numbers of hidden-charm pentaquark states. The formation of diquarks in color antitriplet is supported by the attractive interaction induced by the one-gluon exchange. The QCD sum rules suggest that the scalar and axialvector diquark states are the supported configurations~\cite {Wang:2010sh,Kleiv:2013dta}. Therefore, we have chosen the axialvector-diquark-scalar-diquark-antiquark and the axialvector-diquark-axialvector-diquark-antiquark type interpolating currents for this work, which are provided as follows:

\begin{eqnarray}
J_{\mu}^{1}(x)&=&\frac{\varepsilon_{abc}\varepsilon_{ade} \varepsilon_{bfg}}{\sqrt{3}} \Big[  {q}^T_d(x) C\gamma_\mu {q}_e(x) {q}^T_f(x) C\gamma_5 c_g(x)  C \bar{c}^{T}_{c}(x) \Big] \, , \\
J_{\mu}^{2}(x)&=& \varepsilon_{abc}\varepsilon_{ade} \varepsilon_{bfg}\Big[  {q}^T_d(x) C\gamma_\mu {q}_e(x) {q}^T_f(x) C\gamma_\alpha c_g(x) \gamma_5 \gamma^\alpha C \bar{c}^{T}_{c}(x) \Big] \, ,\\
J_{\mu}^{3}(x)&=& \varepsilon_{abc}\varepsilon_{ade} \varepsilon_{bfg}\Big[  {q}^T_d(x) C\gamma_\alpha {q}_e(x) {q}^T_f(x) C\gamma_\mu c_g(x) \gamma_5 \gamma^\alpha C \bar{c}^{T}_{c}(x) \Big] \, ,
 \end{eqnarray}
where $a$, $b$, $c$, $d$, $e$, $f$ and $g$ being color indices, $q$ stands for u, d or s-quark, and the $C$ is the charge conjugation operator. It is important to note that all of these interpolating currents have identical quantum numbers and quark contents. Therefore, in principle, they should couple to the same pentaquark state in each channel. 

Following this brief introduction, we can now analyze magnetic dipole and higher multipole moments using the QCD light-cone sum rule method. The QCD light-cone sum rule is a well-known and successful technique for calculating the mass, form factor, and magnetic moments of conventional and nonconventional hadron states, as well as elucidating their internal structure~\cite{Chernyak:1990ag, Braun:1988qv, Balitsky:1989ry}. The method calculates the correlation function using two different approaches: the hadronic representation and the QCD representation. The hadronic representation uses hadronic parameters, such as mass, form factors, and so on, while the QCD representation uses QCD parameters such as distribution amplitudes, quark condensate, and so on. Then, to eliminate the effects of the continuum and higher states, we perform the double Borel transformation on both representations of the correlation function concerning the variables $p^2$ and $(p+q)^2$, and apply the quark-hadron duality ansatz.

To compute the hadronic side of the correlation function, after inserting into Eq. (\ref{edmn01}) a full set of states with the spin-parities and contents of the pentaquark states with hidden-charm, and performing integration over x, we obtain
\begin{eqnarray}\label{edmn02}
\Pi^{Had}_{\mu\nu}(p,q)&=&\frac{\langle0\mid J_{\mu}(x)\mid
P_{c}(p_2)\rangle}{[p_2^{2}-m_{P_{c}}^{2}]}\langle P_{c}(p_2)\mid
P_{c}(p_1)\rangle_\gamma\frac{\langle P_{c}(p_1)\mid
\bar{J}_{\nu}(0)\mid 0\rangle}{[p_1^{2}-m_{P_{c}}^{2}]},
\end{eqnarray}
where $p_1 = p+q$, $p_2=p$, and $q$ is the momentum of the photon.  
The matrix elements of the interpolating current between the hidden-charm pentaquark and the vacuum are written in the following way
\begin{align}\label{lambdabey}
\langle0\mid J_{\mu}(x)\mid P_{c}(p_2,s)\rangle &=\lambda_{P_{c}}u_{\mu}(p_2,s),\\
\langle {P_{c}}(p_1,s)\mid
\bar{J}_{\nu}^{P_{c}}(0)\mid 0\rangle &= \lambda_{{P_{c}}}\bar u_{\nu}(p_1,s),
\end{align}
where $\lambda_{P_{c}}$ being the residue of the pentaquark states with hidden-charm, and $u_{\mu}(p,s)$ and $ \bar u_{\nu}(p+q,s)$ are the Rarita-Schwinger spinors. 
 
The matrix element of the radiative transition in the Eq.(\ref{edmn02}) can be described in the following way\cite{Weber:1978dh,Nozawa:1990gt,Pascalutsa:2006up,Ramalho:2009vc}:
\begin{eqnarray}\label{matelpar}
\langle P_{c}(p_2)\mid P_{c}(p_1)\rangle_\gamma &=&-e\bar
u_{\mu}(p_2)\left\{F_{1}(q^2)g_{\mu\nu}\eslash-
\frac{1}{2m_{P_{c}}}\left
[F_{2}(q^2)g_{\mu\nu}+F_{4}(q^2)\frac{q_{\mu}q_{\nu}}{(2m_{P_{c}})^2}\right]\eslash\qslash
\right.\nonumber\\&+&\left.
F_{3}(q^2)\frac{1}{(2m_{P_{c}})^2}q_{\mu}q_{\nu}\eslash\right\} u_{\nu}(p_1),
\end{eqnarray}
where $\varepsilon$ denotes the  polarization vector of the photon, and $\rm{F_i}$ are the Lorentz invariant form factors.  The correlation function concerning hadronic observables is achieved by utilizing the Eqs. (\ref{edmn02})-(\ref{matelpar}). After completing these steps, the final version of this function is  
\begin{align}\label{fizson}
 \Pi^{Had}_{\mu\nu}(p,q)&=-\frac{\lambda_{_{P_{c}}}^{2}\,\big(\pslash_1+m_{P_{c}}\big)}{[p_1^{2}-m_{_{P_{c}}}^{2}][p_2^{2}-m_{_{P_{c}}}^{2}]}
 \bigg[g_{\mu\nu}
-\frac{1}{3}\gamma_{\mu}\gamma_{\nu}-\frac{2\,p_{1\mu}p_{1\nu}}
{3\,m^{2}_{P_{c}}}+\frac{p_{1\mu}\gamma_{\nu}-p_{1\nu}\gamma_{\mu}}{3\,m_{P_{c}}}\bigg] \bigg\{F_{1}(q^2)g_{\mu\nu}\eslash  -
\frac{1}{2m_{P_{c}}}
\Big[F_{2}(q^2)g_{\mu\nu}\nonumber\\
&  +F_{4}(q^2) \frac{q_{\mu}q_{\nu}}{(2m_{P_{c}})^2}\Big]\eslash\qslash+\frac{F_{3}(q^2)}{(2m_{P_{c}})^2}
 q_{\mu}q_{\nu}\eslash\bigg\}
 \big(\pslash_2+m_{P_{c}}\big)
 \bigg[g_{\mu\nu}-\frac{1}{3}\gamma_{\mu}\gamma_{\nu}-\frac{2\,p_{2\mu}p_{2\nu}}
{3\,m^{2}_{P_{c}}}+\frac{p_{2\mu}\gamma_{\nu}-p_{2\nu}\gamma_{\mu}}{3\,m_{P_{c}}}\bigg].
\end{align}
Here, a summation of the spins of the hidden-charm pentaquark has also been carried out as:
\begin{align}\label{raritabela}
\sum_{s}u_{\mu}(p,s)\bar u_{\nu}(p,s)=-\big(\pslash+m_{P_{c}}\big)\Big[g_{\mu\nu}
-\frac{1}{3}\gamma_{\mu}\gamma_{\nu}-\frac{2\,p_{\mu}p_{\nu}}
{3\,m^{2}_{P_{c}}}+\frac{p_{\mu}\gamma_{\nu}-p_{\nu}\gamma_{\mu}}{3\,m_{P_{c}}}\Big].
\end{align}

In principle, one could use the above equations to obtain the result of the hadronic representation of the analysis, but at this point, we run into two problems that we have to solve and that are likely to affect the reliability of the calculations.  The first one is the non-independence of all Lorentz structures in the correlation function, and the second one is the effects of spin-1/2 pentaquark states, which should also be excluded from the correlation function. The matrix element of the current $J_{\mu}$ between the spin-1/2 pentaquark states and the vacuum is written the following formula
\begin{equation}\label{spin12}
\langle0\mid J_{\mu}(0)\mid B(p,s=1/2)\rangle=(A  p_{\mu}+B\gamma_{\mu})u(p,s=1/2).
\end{equation}
As can be seen, the undesirable contributions associated with the spin-1/2 pentaquark states are related to the $p_\mu$ and the $\gamma_\mu$.  
To get rid of the contaminations coming from the spin-1/2 states and to get only independent structures in the correlation function, we perform the ordering for Dirac matrices as $\gamma_{\mu}\pslash\eslash\qslash\gamma_{\nu}$ and remove terms with $\gamma_\mu$ at the beginning, $\gamma_\nu$ at the end or those proportional to $p_{2\mu}$ or $p_{1\nu}$~\cite{Belyaev:1982cd}. Thus, with the help of the Eqs. (\ref{edmn02})-(\ref{matelpar}) for the hadronic representation, one obtains
\begin{eqnarray}\label{final phenpart}
\Pi^{Had}_{\mu\nu}(p,q)&=&-\frac{\lambda_{_{P_{c}}}^{2}}{[(p+q)^{2}-m_{_{P_{c}}}^{2}][p^{2}-m_{_{P_{c}}}^{2}]}
\bigg[  -g_{\mu\nu}\pslash\eslash\qslash \,F_{1}(q^2) 
+m_{P_{c}}g_{\mu\nu}\eslash\qslash\,F_{2}(q^2)+
\frac{F_{3}(q^2)}{4m_{P_{c}}}q_{\mu}q_{\nu}\eslash\qslash\, \nonumber\\&+&
\frac{F_{4}(q^2)}{4m_{P_{c}}^3}(\varepsilon.p)q_{\mu}q_{\nu}\pslash\qslash \,+
\rm{other~independent~structures} \bigg].
\end{eqnarray}

Using the form factors $F_{i}(q^2)$, the magnetic dipole, $G_{M}(q^2)$, electric quadrupole, $G_{Q}(q^2)$, and magnetic octupole, $G_{O}(q^2)$, form factors are described by the following equations~\cite{Weber:1978dh,Nozawa:1990gt,Pascalutsa:2006up,Ramalho:2009vc}:
\begin{eqnarray}
G_{M}(q^2) &=& \left[ F_1(q^2) + F_2(q^2)\right] ( 1+ \frac{4}{5}
\tau ) -\frac{2}{5} \left[ F_3(q^2)  +
F_4(q^2)\right] \tau \left( 1 + \tau \right), \nonumber\\
G_{Q}(q^2) &=& \left[ F_1(q^2) -\tau F_2(q^2) \right]  -
\frac{1}{2}\left[ F_3(q^2) -\tau F_4(q^2)
\right] \left( 1+ \tau \right),  \nonumber \\
 G_{O}(q^2) &=&
\left[ F_1(q^2) + F_2(q^2)\right] -\frac{1}{2} \left[ F_3(q^2)  +
F_4(q^2)\right] \left( 1 + \tau \right),\end{eqnarray}
  where $\tau
= -\frac{q^2}{4m^2_{P_{c}}}$. In the static limit, i.e. $q^2=0$, the magnetic dipole and higher multipole form factors with respect to the functions $F_i(0)$ are given as
\begin{eqnarray}\label{mqo1}
G_{M}(0)&=&F_{1}(0)+F_{2}(0),\nonumber\\
G_{Q}(0)&=&F_{1}(0)-\frac{1}{2}F_{3}(0),\nonumber\\
G_{O}(0)&=&F_{1}(0)+F_{2}(0)-\frac{1}{2}[F_{3}(0)+F_{4}(0)].
\end{eqnarray}
From the above equations, the  magnetic dipole ($\mu_{P_{c}}$), the electric quadrupole
($Q_{P_{c}}$)  and the magnetic octupole ($O_{P_{c}}$) moments are obtained as 
 \begin{eqnarray}\label{mqo2}
\mu_{P_{c}}&=&\frac{e}{2m_{P_{c}}}G_{M}(0),~~~~~~
Q_{P_{c}}=\frac{e}{m_{P_{c}}^2}G_{Q}(0), ~~~~~~
O_{P_{c}}=\frac{e}{2m_{P_{c}}^3}G_{O}(0).
\end{eqnarray}

It is important to mention that it is possible to have more than one hidden-charm pentaquark with the specified quark content that can couple to a given interpolating current. If the hidden-charm pentaquarks have substantially different masses, then Eqs. (\ref{edmn02})-(\ref{mqo2}) can be applied to the lowest mass hidden-charm pentaquark, and in this scenario, the magnetic dipole and higher multipole moments extracted with different interpolating currents should yield the same value. However, If the masses of these hidden-charm pentaquarks are similar to one another, it means that they are close to degenerate states. In that scenario, Eq. (\ref{mqo2}) provides a weighted average where the weighing computed by the residues squares, of the magnetic dipole and higher multipole moments of these hidden-charm pentaquarks. It should be noted that the magnetic dipole and higher multipole moments acquired by utilizing different currents may differ if the magnetic dipole and higher multipole moments of these nearly degenerate hidden-charm pentaquark states differ significantly. This is due to the fact that the magnetic dipole and higher multipole moments are extracted by using different weight factors for different currents.

The QCD side of the sum rules is formed by the correlation function, which is expressed using quark propagators and evaluated with some accuracy in the operator product expansion. When the quark fields are contracted, the correlation function becomes an expression that is given regarding the quark propagators, as in the following equations
\begin{eqnarray}
\label{QCD1}
\Pi^{QCD-J_\mu^1}_{\mu\nu}(p,q)&=-&\frac{i}{3}\,\mathcal{A} \int d^4x e^{ip\cdot x} \langle 0|
\Big\{ \mbox{Tr}\Big[\gamma_\mu S_q^{ee^\prime}(x) \gamma_\nu \tilde S_q^{dd^\prime}(x)\Big]
 \mbox{Tr}\Big[\gamma_5 S_c^{gg^\prime}(x) \gamma_5 \tilde S_q^{ff^\prime}(x)\Big] 
 \nonumber\\
&& -   \mbox{Tr} \Big[\gamma_\mu S_q^{ef^\prime}(x) \gamma_5 \tilde S_c^{gg^\prime}(x) 
\gamma_5 S_q^{fe^\prime}(x) \gamma_\nu \tilde S_q^{dd^\prime}(x)\Big]
-
 \mbox{Tr} \Big[\gamma_\mu S_q^{ee^\prime}(x) \gamma_\nu \tilde S_q^{fd^\prime}(x) 
\gamma_5 S_c^{gg^\prime}(x) \gamma_5 \tilde S_q^{df^\prime}(x)\Big]
\nonumber\\
&& +
 \mbox{Tr} \Big[\gamma_\mu S_q^{ed^\prime}(x) \gamma_\nu \tilde S_q^{fe^\prime}(x) 
\gamma_5 S_c^{gg^\prime}(x) \gamma_5 \tilde S_q^{df^\prime}(x)\Big]
+
 \mbox{Tr} \Big[\gamma_\mu S_q^{ef^\prime}(x) \gamma_\nu \tilde S_c^{gg^\prime}(x) 
\gamma_5 S_q^{fd^\prime}(x) \gamma_5 \tilde S_q^{de^\prime}(x)\Big]
\nonumber\\
&&-
\mbox{Tr}\Big[\gamma_\mu S_q^{ed^\prime}(x) \gamma_\nu \tilde S_q^{de^\prime}(x)\Big]
 \mbox{Tr}\Big[\gamma_5 S_c^{gg^\prime}(x) \gamma_5 \tilde S_q^{ff^\prime}(x)\Big] 
\Big \}  \tilde S_c^{c^{\prime}c}(-x)
|0 \rangle_\gamma ,\\
\nonumber\\
%
\label{QCD2}
\Pi^{QCD-J_\mu^2}_{\mu\nu}(p,q)&=& i\,\mathcal{A} \int d^4x e^{ip\cdot x} \langle 0|
\Big\{ \mbox{Tr}\Big[\gamma_\mu S_q^{ee^\prime}(x) \gamma_\nu \tilde S_q^{dd^\prime}(x)\Big]
 \mbox{Tr}\Big[\gamma_\alpha S_c^{gg^\prime}(x) \gamma_\beta \tilde S_q^{ff^\prime}(x)\Big] 
 \nonumber\\
&& -   \mbox{Tr} \Big[\gamma_\mu S_q^{ef^\prime}(x) \gamma_\alpha \tilde S_c^{gg^\prime}(x) 
\gamma_\beta S_q^{fe^\prime}(x) \gamma_\nu \tilde S_q^{dd^\prime}(x)\Big]
-
 \mbox{Tr} \Big[\gamma_\mu S_q^{ee^\prime}(x) \gamma_\nu \tilde S_q^{fd^\prime}(x) 
\gamma_\alpha S_c^{gg^\prime}(x) \gamma_\beta \tilde S_q^{df^\prime}(x)\Big]
\nonumber\\
&& +
 \mbox{Tr} \Big[\gamma_\mu S_q^{ed^\prime}(x) \gamma_\nu \tilde S_q^{fe^\prime}(x) 
\gamma_\alpha S_c^{gg^\prime}(x) \gamma_\beta \tilde S_q^{df^\prime}(x)\Big]
+
 \mbox{Tr} \Big[\gamma_\mu S_q^{ef^\prime}(x) \gamma_\nu \tilde S_c^{gg^\prime}(x) 
\gamma_\alpha S_q^{fd^\prime}(x) \gamma_\beta \tilde S_q^{de^\prime}(x)\Big]
\nonumber\\
&&-
\mbox{Tr}\Big[\gamma_\mu S_q^{ed^\prime}(x) \gamma_\nu \tilde S_q^{de^\prime}(x)\Big]
 \mbox{Tr}\Big[\gamma_\alpha S_c^{gg^\prime}(x) \gamma_\beta \tilde S_q^{ff^\prime}(x)\Big] 
\Big \} \Big(\gamma_5 \gamma^\alpha \tilde S_c^{c^{\prime}c}(-x) \gamma^\beta \gamma_5 \Big)
|0 \rangle_\gamma ,
\\
\nonumber\\
 \label{QCD3}
 \Pi^{QCD-J_\mu^3}_{\mu\nu}(p,q)&=&i\,\mathcal{A} \int d^4x e^{ip\cdot x} \langle 0|
\Big\{ \mbox{Tr}\Big[\gamma_\alpha S_q^{ee^\prime}(x) \gamma_\beta \tilde S_q^{dd^\prime}(x)\Big]
 \mbox{Tr}\Big[\gamma_\mu S_c^{gg^\prime}(x) \gamma_\nu \tilde S_q^{ff^\prime}(x)\Big] 
 \nonumber\\
&& -   \mbox{Tr} \Big[\gamma_\alpha S_q^{ef^\prime}(x) \gamma_\mu \tilde S_c^{gg^\prime}(x) 
\gamma_\nu S_q^{fe^\prime}(x) \gamma_\beta \tilde S_q^{dd^\prime}(x)\Big]
-
 \mbox{Tr} \Big[\gamma_\alpha S_q^{ee^\prime}(x) \gamma_\beta \tilde S_q^{fd^\prime}(x) 
\gamma_\mu S_c^{gg^\prime}(x) \gamma_\nu \tilde S_q^{df^\prime}(x)\Big]
\nonumber\\
&& +
 \mbox{Tr} \Big[\gamma_\alpha S_q^{ed^\prime}(x) \gamma_\beta \tilde S_q^{fe^\prime}(x) 
\gamma_\mu S_c^{gg^\prime}(x) \gamma_\nu \tilde S_q^{df^\prime}(x)\Big]
+
 \mbox{Tr} \Big[\gamma_\alpha S_q^{ef^\prime}(x) \gamma_\beta \tilde S_c^{gg^\prime}(x) 
\gamma_\mu S_q^{fd^\prime}(x) \gamma_\nu \tilde S_q^{de^\prime}(x)\Big]
\nonumber\\
&&-
\mbox{Tr}\Big[\gamma_\alpha S_q^{ed^\prime}(x) \gamma_\beta \tilde S_q^{de^\prime}(x)\Big]
 \mbox{Tr}\Big[\gamma_\mu S_c^{gg^\prime}(x) \gamma_\nu \tilde S_q^{ff^\prime}(x)\Big] 
\Big \} \Big(\gamma_5 \gamma^\alpha \tilde S_c^{c^{\prime}c}(-x) \gamma^\beta \gamma_5 \Big)
|0 \rangle_\gamma ,
\end{eqnarray}
 where $\mathcal{A}= \varepsilon^{abc}\varepsilon^{a^{\prime}b^{\prime}c^{\prime}}\varepsilon^{ade}
\varepsilon^{a^{\prime}d^{\prime}e^{\prime}}\varepsilon^{bfg}
\varepsilon^{b^{\prime}f^{\prime}g^{\prime}}$ and 
$ \widetilde{S}_{c(q)}^{ij}(x)=CS_{c(q)}^{ij\mathrm{T}}(x)C$,  %
with $S_{q}(x)$ and $S_{c}(x)$ are the quark propagators, and the explicit forms of these propagators being~\cite{Yang:1993bp, Belyaev:1985wza}
\begin{align}
\label{edmn12}
S_{q}(x)&= \frac{1}{2 \pi x^2}\Big(i \frac{\xslash}{x^2}- \frac{m_q}{2}\Big) 
- \frac{\langle \bar qq \rangle }{12} \Big(1-i\frac{m_{q} \xslash}{4}   \Big)
- \frac{ \langle \bar qq \rangle }{192}
m_0^2 x^2  \Big(1-i\frac{m_{q} \xslash}{6}   \Big)
-\frac {i g_s }{32 \pi^2 x^2} ~G^{\mu \nu} (x) 
\Big[\rlap/{x} 
\sigma_{\mu \nu} +  \sigma_{\mu \nu} \rlap/{x}
 \Big],
\end{align}%
%
\begin{align}
\label{edmn13}
S_{c}(x)&=\frac{m_{c}^{2}}{4 \pi^{2}} \bigg[ \frac{K_{1}\Big(m_{c}\sqrt{-x^{2}}\Big) }{\sqrt{-x^{2}}}
+i\frac{{\xslash}~K_{2}\Big( m_{c}\sqrt{-x^{2}}\Big)}
{(\sqrt{-x^{2}})^{2}}\Bigg]
-\frac{g_{s}m_{c}}{16\pi ^{2}} \int_0^1 dv\, G^{\mu \nu }(vx)\Bigg[ (\sigma _{\mu \nu }{\xslash}
  +{\xslash}\sigma _{\mu \nu })\frac{K_{1}\Big( m_{c}\sqrt{-x^{2}}\Big) }{\sqrt{-x^{2}}}\nonumber\\
&+2\sigma_{\mu \nu }K_{0}\Big( m_{c}\sqrt{-x^{2}}\Big)\bigg],
\end{align}%
where $G^{\mu\nu}$ is the gluon field strength tensor and $K_i$ are modified the second kind Bessel functions.

Two different contributions, perturbative and non-perturbative, are included in the correlation functions obtained on the QCD representation. The perturbative effects are associated with the perturbative soft interaction of the photon with the light and heavy quarks.  To evaluate such effects, one of the quark propagators, interacting with the photon is substituted through
\begin{align}
\label{free}
S^{free}(x) \rightarrow \int d^4y\, S^{free} (x-y)\,\rlap/{\!A}(y)\, S^{free} (y)\,,
\end{align}
with $S^{free}(x)$ stands for the first term of the quark propagators and in Eqs. (\ref{QCD1})-(\ref{QCD3}) the surviving four propagators are being substituted by full quark propagators.

We use the following replacement for one of the light quark propagators interacting with the photon to obtain the non-perturbative effects representing the interaction of the photon with light quark lines on a large scale
\begin{align}
\label{edmn14}
S_{\alpha\beta}^{ab}(x) \rightarrow -\frac{1}{4} \big[\bar{q}^a(x) \Gamma_i q^b(0)\big]\big(\Gamma_i\big)_{\alpha\beta},
\end{align}
where   $\Gamma_i = \mathrm{1}, \gamma_5, \gamma_\mu, i\gamma_5 \gamma_\mu, \sigma_{\mu\nu}/2$. After the light-quark replacement mentioned above, the remaining propagators are considered to be full propagators. 
The matrix elements of nonlocal operators such as $\langle \gamma(q)\vel \bar{q}(x) \Gamma_i G_{\mu\nu}q(0) \ver 0\rangle$  and $\langle \gamma(q)\vel \bar{q}(x) \Gamma_i q(0) \ver 0\rangle$, expressed concerning photon distribution amplitudes (DAs), which occur once a photon interacts non-perturbatively with light-quark fields (for details see Ref. \cite{Ball:2002ps}).  The correlation function concerning QCD parameters is extracted by means of the above procedures and then by applying the Fourier transform to the obtained expressions in order to transfer the x-space expressions to the momentum space.

The QCD light-cone sum rules for the  $F_1$, $F_2$, $F_3$, and $F_4$ form factors can be determined by equating the coefficients of the $g_{\mu\nu}\pslash\eslash\qslash$, $g_{\mu\nu}\eslash\qslash$,
$q_{\mu}q_{\nu}\eslash\qslash$ and  
$(\varepsilon.p)q_{\mu}q_{\nu}\pslash\qslash$ structures.  
Since all the results are more or less similar, for simplicity only the results of the $J^1_{\mu}$  interpolating current are listed in the Appendix.

\end{widetext}

\section{Numerical analysis and discussion}\label{numerical}
In this section, we perform the QCD light-cone sum rule numerical analyses to predict the magnetic dipole and higher multipole moments hidden-charm pentaquark states. 
 There are various parameters for which we require their numerical values for these physical observables in the QCD light-cone sum rule. 
We take  $m_u =m_d =0$, 
$m_s =93.4^{+8.6}_{-3.4}\,\mbox{MeV}$, $m_c = 1.27 \pm 0.02\,\mbox{GeV}$~\cite{ParticleDataGroup:2022pth},  
$m_{P_{c\bar c uuu}}^{J_\mu^1} = 4.39 \pm 0.13$ GeV,  
$m_{P_{c\bar c ddd}}^{J_\mu^1} = 4.39 \pm 0.13$ GeV,  
$m_{P_{c\bar c sss}}^{J_\mu^1} = 4.70 \pm 0.11$ GeV, 
$m_{P_{c\bar c uuu}}^{J_\mu^2} = 4.39 \pm 0.14$ GeV,  
$m_{P_{c\bar c ddd}}^{J_\mu^2} = 4.39 \pm 0.14$ GeV,  
$m_{P_{c\bar c sss}}^{J_\mu^2} = 4.71 \pm 0.11$ GeV, 
$m_{P_{c\bar c uuu}}^{J_\mu^3} = 4.39 \pm 0.14$ GeV,  
$m_{P_{c\bar c ddd}}^{J_\mu^3} = 4.39 \pm 0.14$ GeV,  
$m_{P_{c\bar c sss}}^{J_\mu^3} = 4.72 \pm 0.11$ GeV~\cite{Wang:2015wsa}, 
$\langle \bar uu\rangle = 
\langle \bar dd\rangle=(-0.24 \pm 0.01)^3\,\mbox{GeV}^3$, $\langle \bar ss\rangle = 0.8\, \langle \bar uu\rangle$ $\,\mbox{GeV}^3$ \cite{Ioffe:2005ym},
$m_0^{2} = 0.8 \pm 0.1 \,\mbox{GeV}^2$ \cite{Ioffe:2005ym},  and 
$\langle g_s^2G^2\rangle = 0.48 \pm 0.14~ \mbox{GeV}^4$~\cite{Narison:2018nbv}.   For further calculations, we require the residues of these hidden-charm pentaquarks, which have been taken from Ref.~\cite{Wang:2015wsa}.

\begin{widetext}
 
 \begin{figure}[htp]
\centering
\subfloat[]{\includegraphics[width=0.30\textwidth]{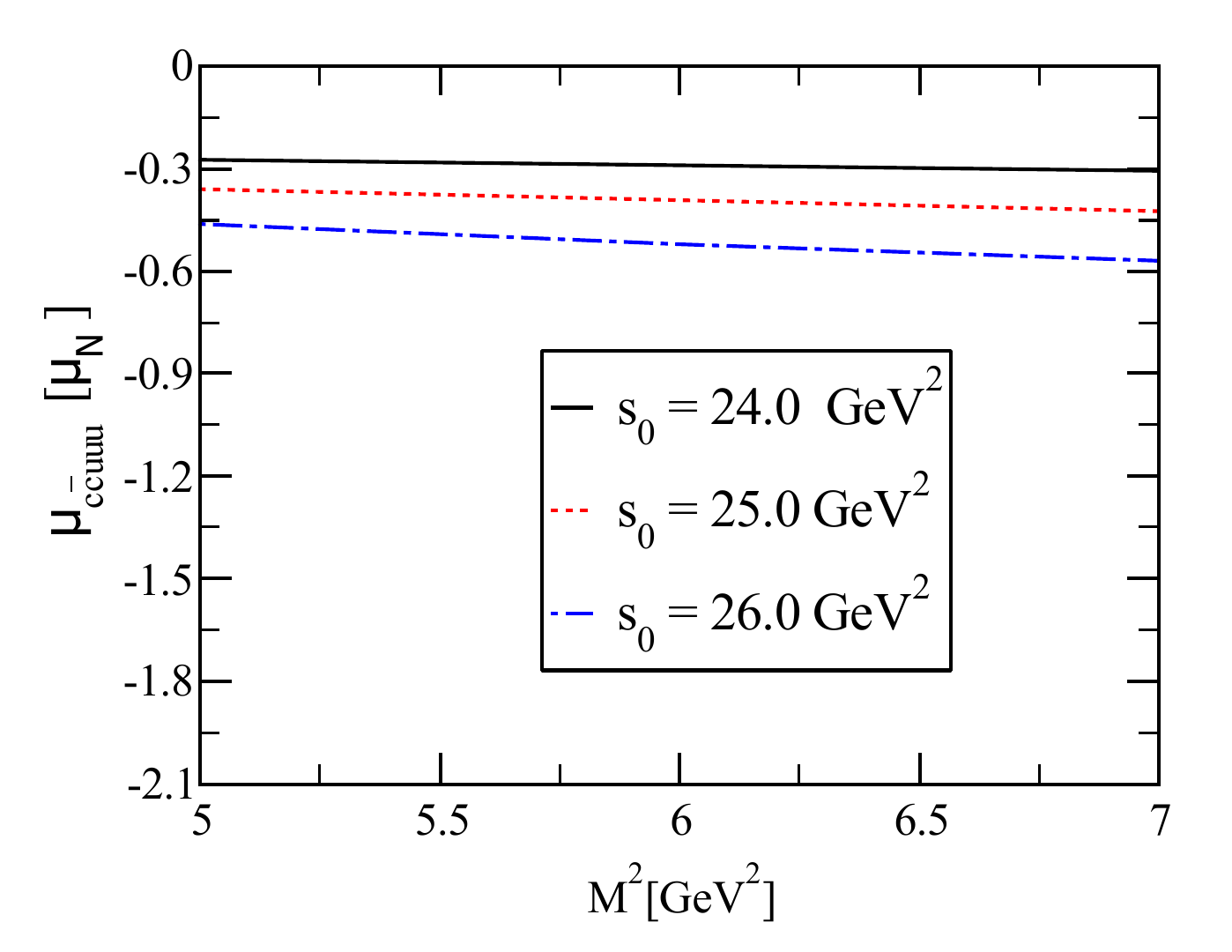}}
\subfloat[]{\includegraphics[width=0.30\textwidth]{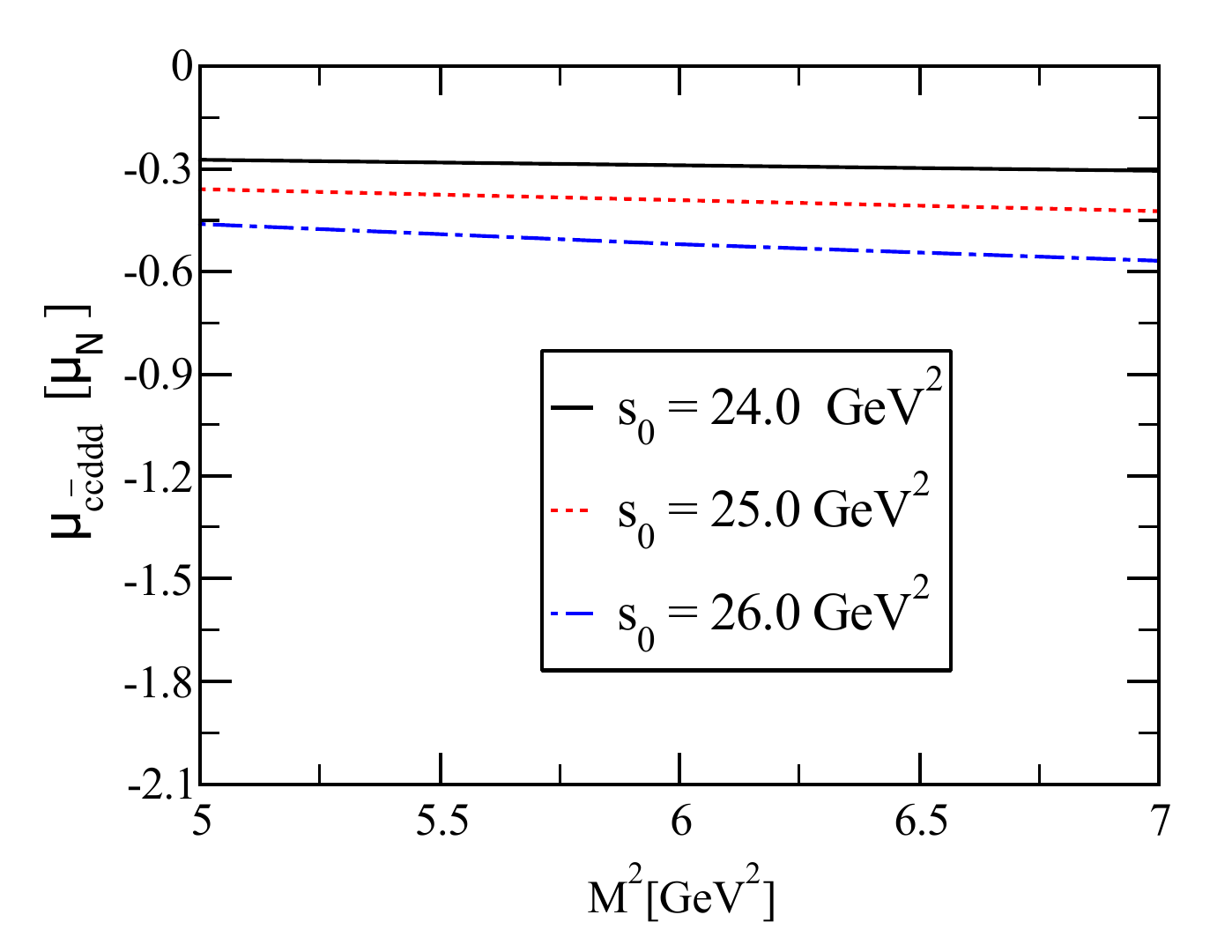}}
\subfloat[]{\includegraphics[width=0.30\textwidth]{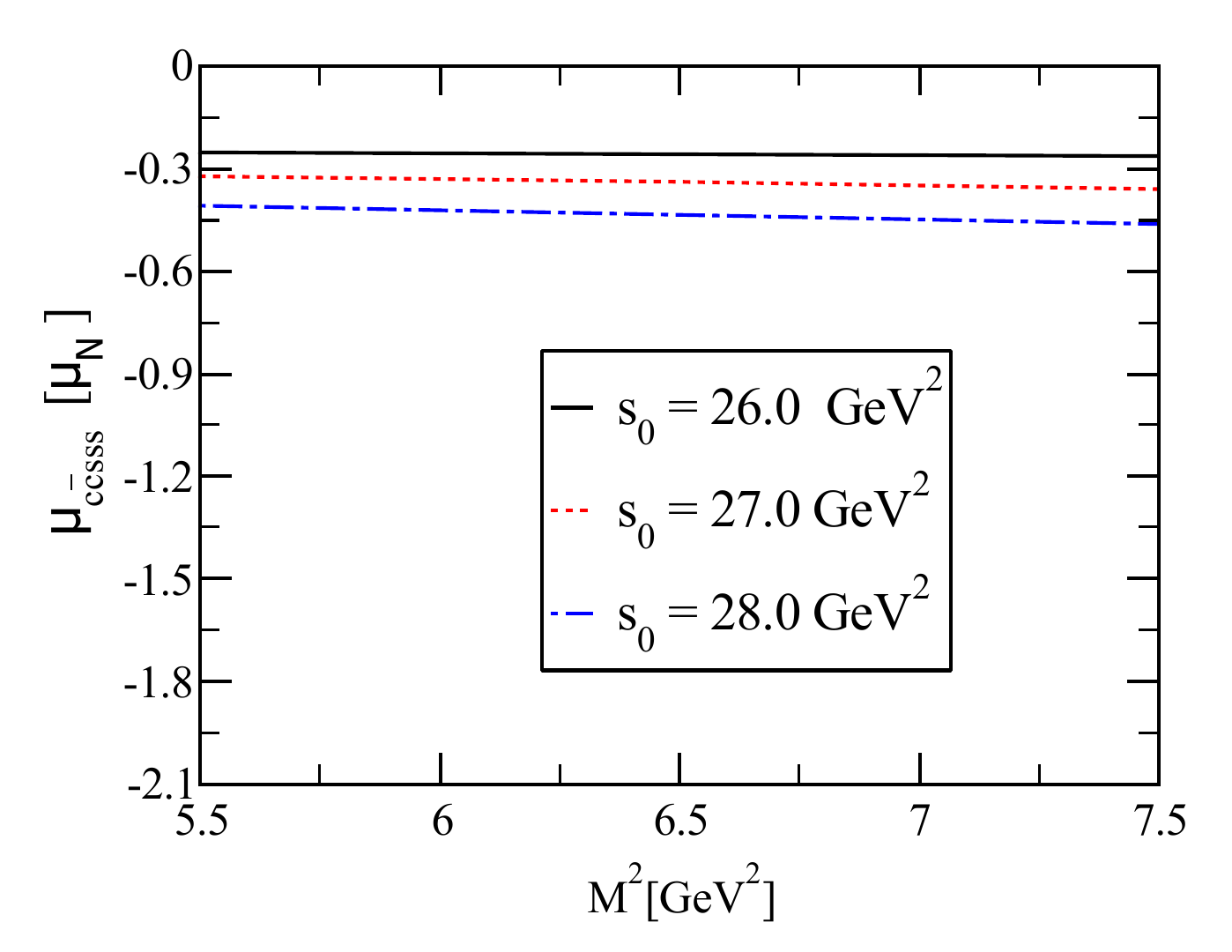}}\\
\subfloat[]{\includegraphics[width=0.30\textwidth]{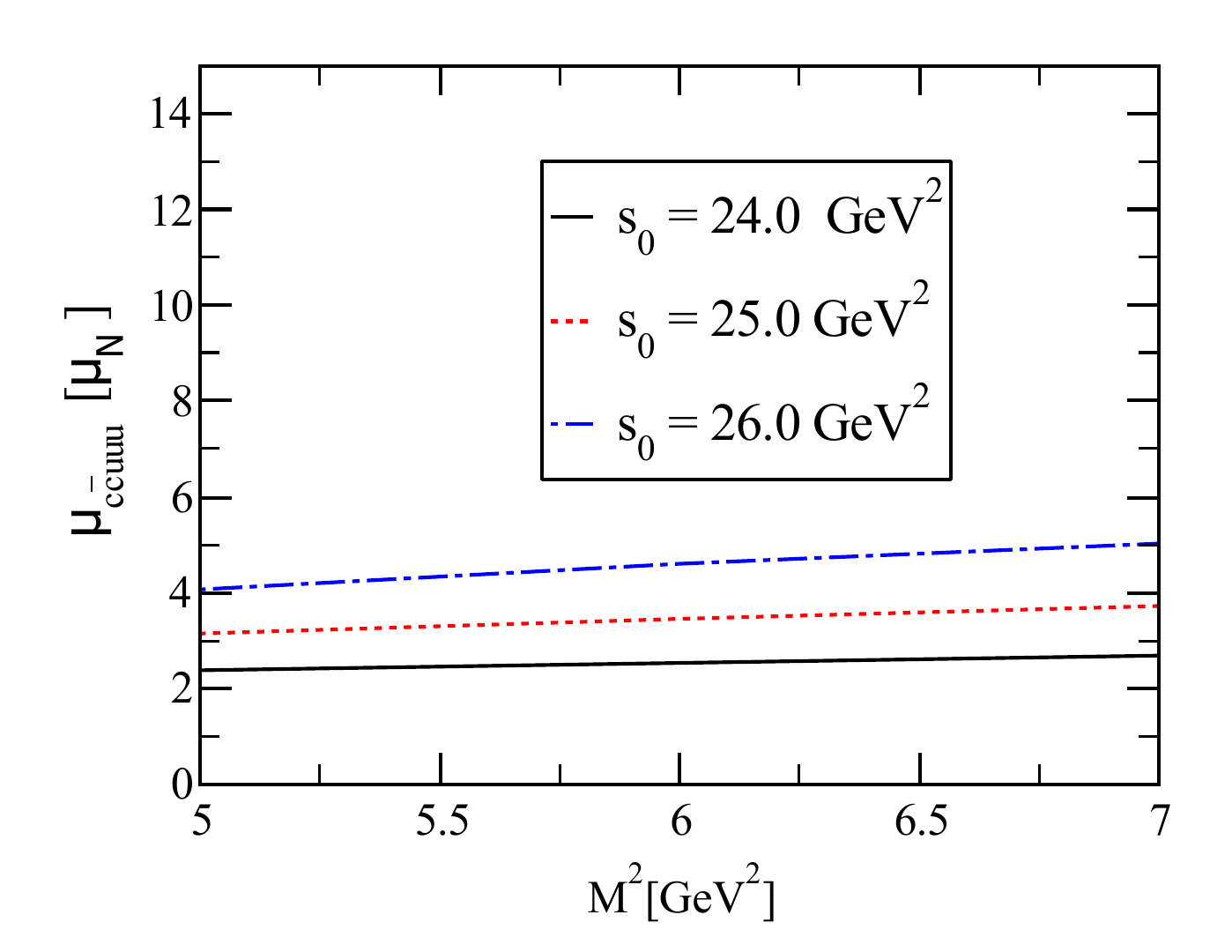}}
\subfloat[]{\includegraphics[width=0.30\textwidth]{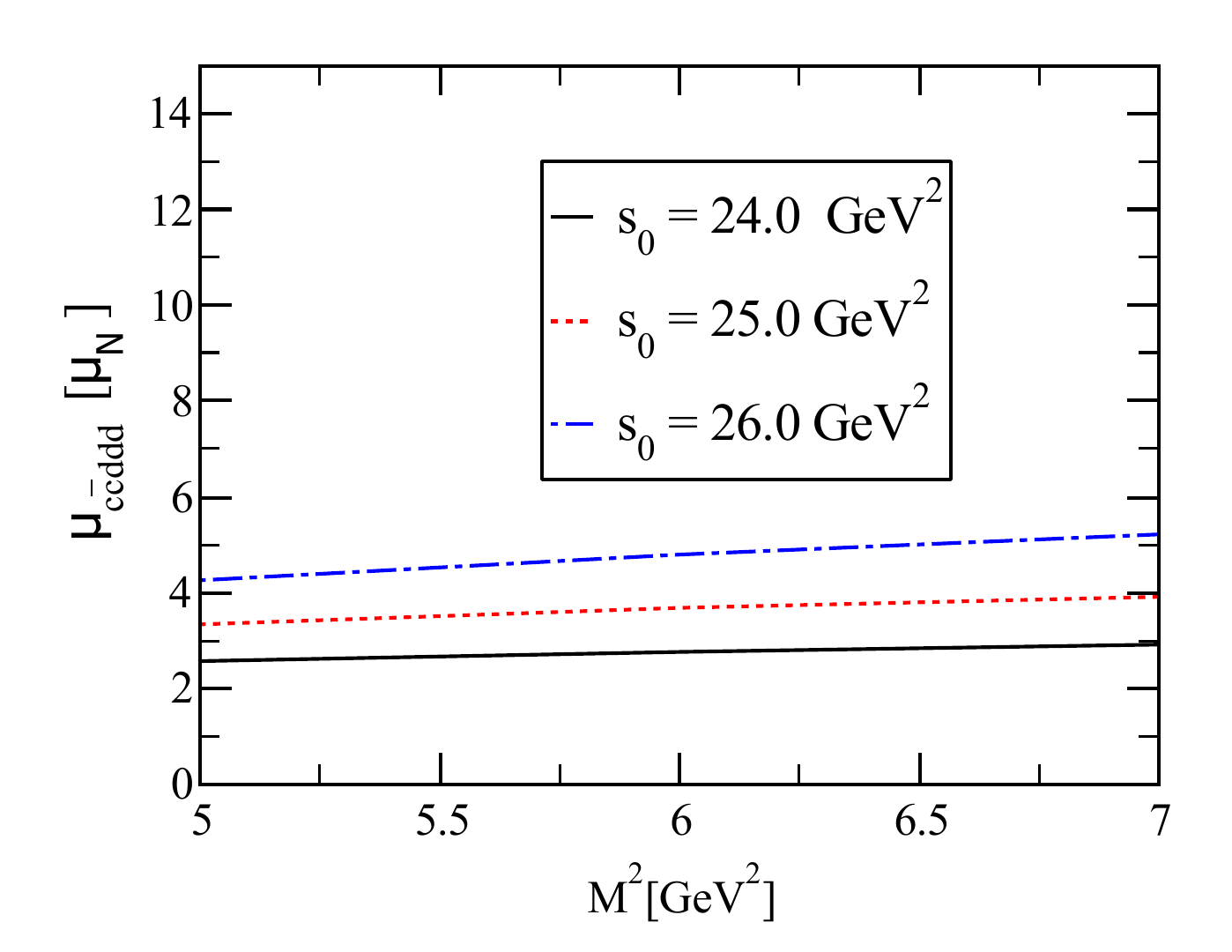}}
\subfloat[]{\includegraphics[width=0.30\textwidth]{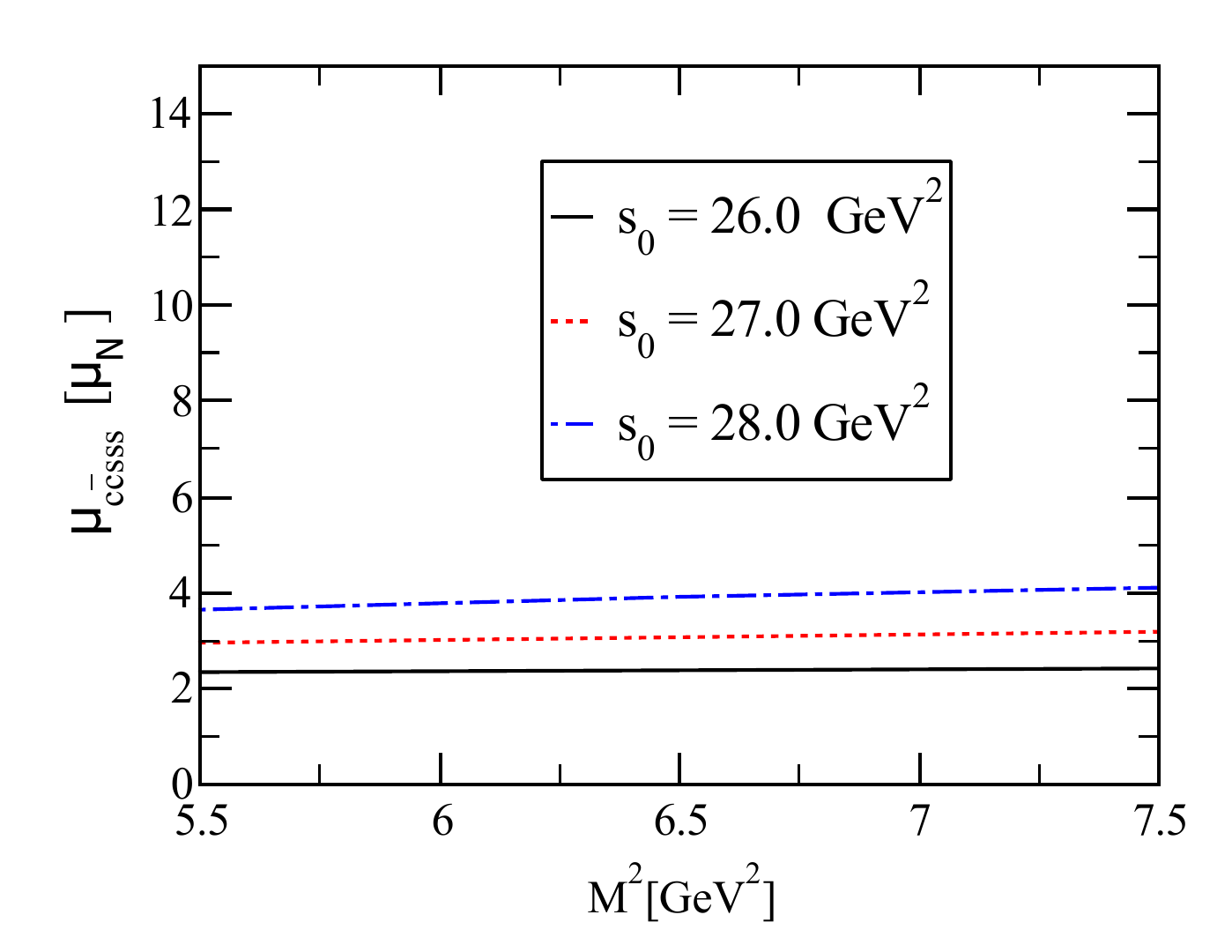}}\\
\subfloat[]{\includegraphics[width=0.30\textwidth]{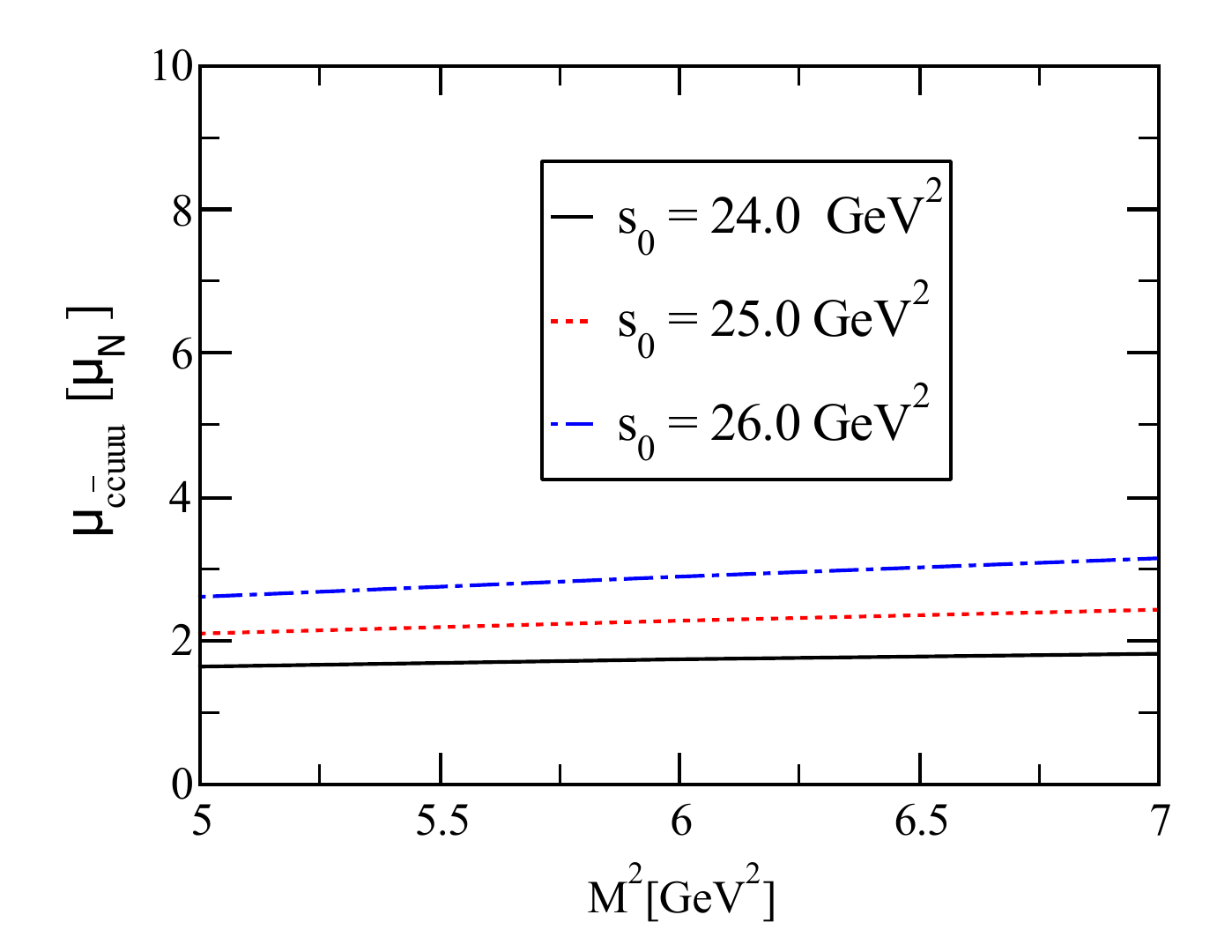}}
\subfloat[]{\includegraphics[width=0.30\textwidth]{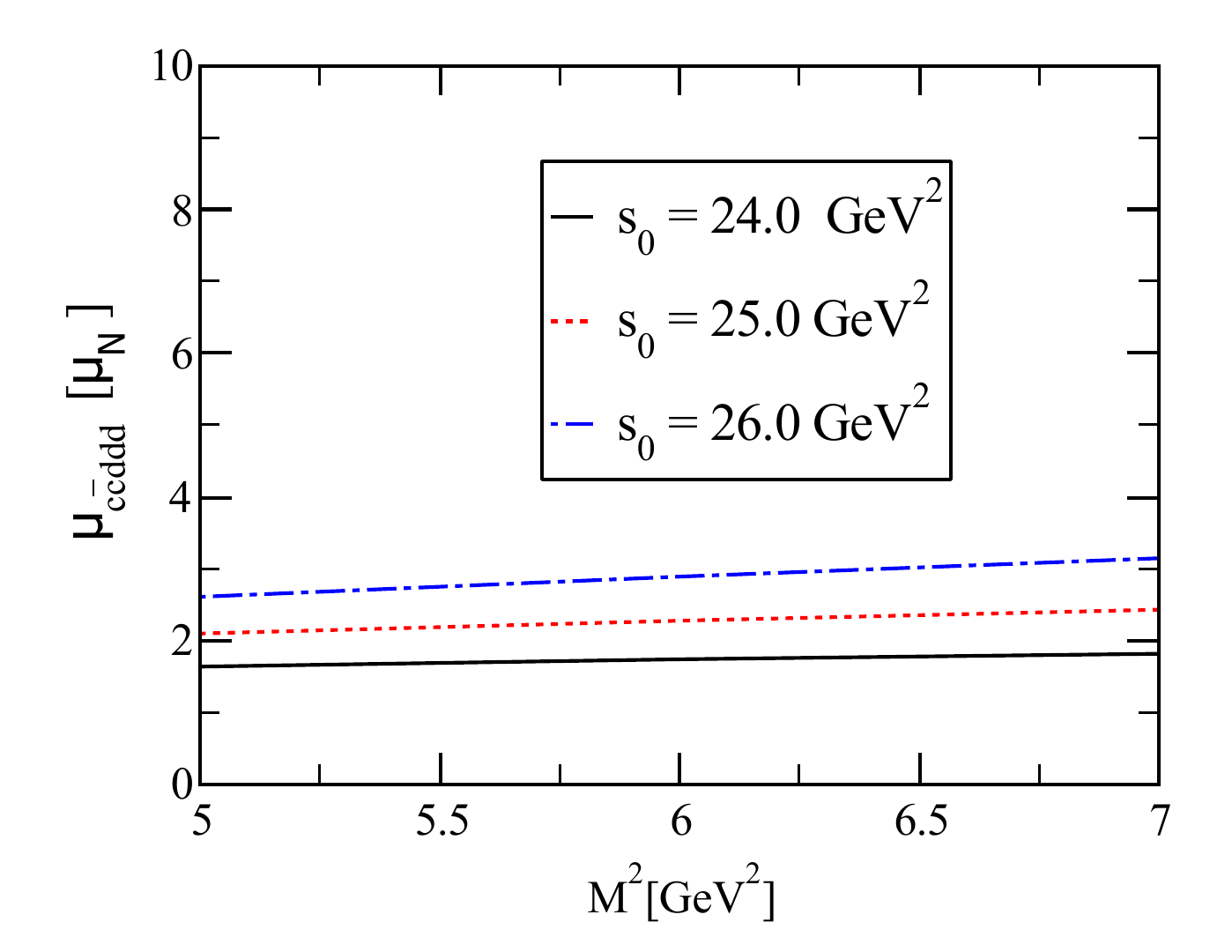}}
\subfloat[]{\includegraphics[width=0.30\textwidth]{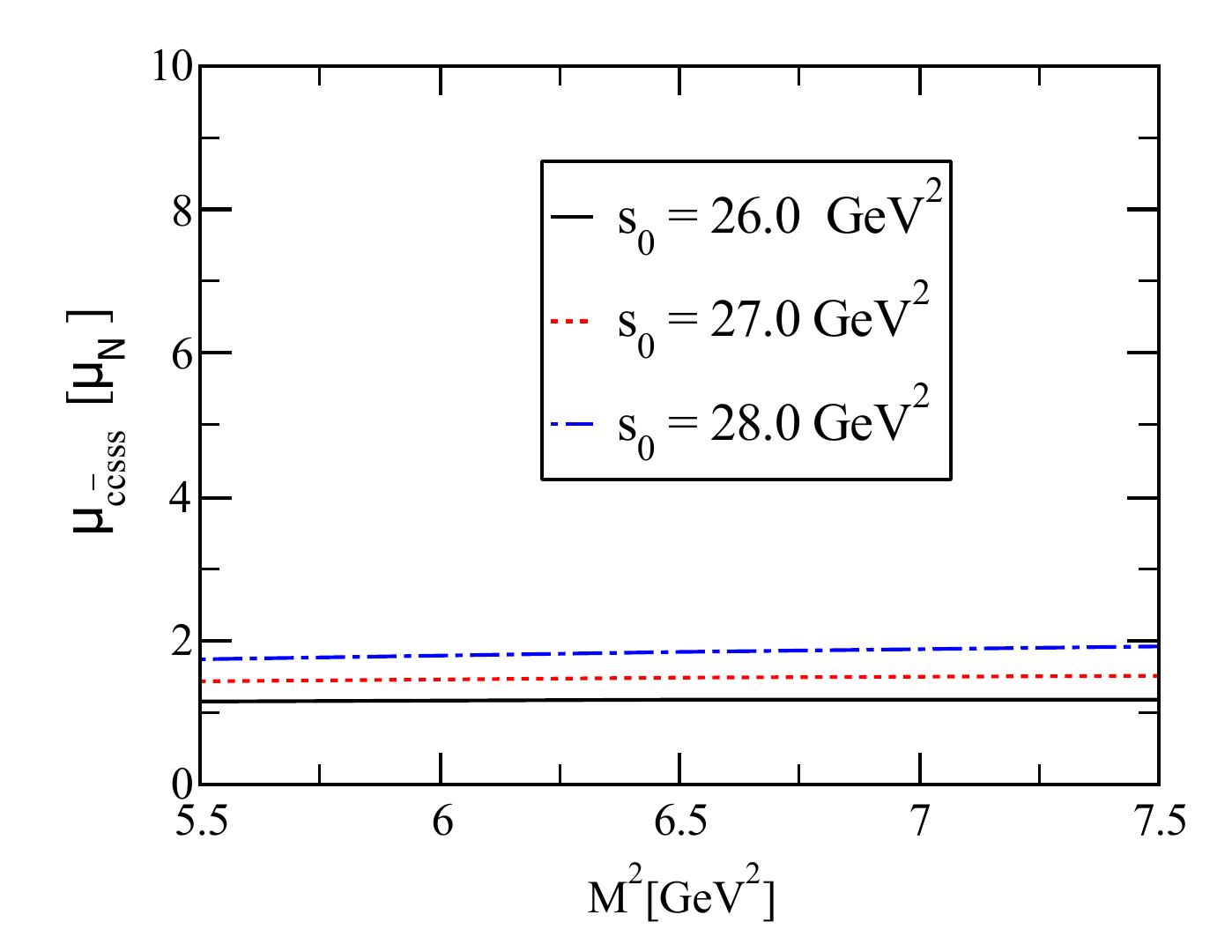}}\\
 \caption{The magnetic dipole moments of the hidden-charm pentaquark states versus $\rm{M^2}$ for fixed values of $\rm{s_0}$; (a), (b)  and (c) for $J^1_\mu$ current, (d), (e) and (f) for $J^2_\mu$ current, and; (g), (h) and (i) for $J^3_\mu$ current. 
 }
 \label{Msqfig}
  \end{figure}

  \end{widetext}

There are two helping parameters in addition to the ones listed above: the Borel parameter $\rm{M^2}$ and the threshold parameter $\rm{s_0}$. The working regions of these auxiliary parameters, where the results show weak dependence on these parameters, are determined by analyzing the results based on the QCD sum rule technique. The $\rm{s_0}$ is not entirely arbitrary. It is the point at which the correlation function begins to include contributions from both excited states and the continuum. 
To determine this parameter, it is generally accepted to assume that $\rm{s_0} = (M_H + 0.5^{+0.1}_{-0.1}) \rm{GeV }^2$. The results are then analyzed for their dependence on slight variations of this parameter. The upper and lower values of the $\rm{M^2}$ are determined by the pole dominance (PC) and the convergence of the operator product expansion (CVG). The CVG generally needs to be small enough to ensure convergence of the operator product expansion series and the PC needs to be as large as possible to guarantee the validity of the one-pole approximation. These constraints can be described using the following equations:
\begin{align}
 \rm{PC} &=\frac{\Delta (\rm{M^2},\rm{s_0})}{\Delta (\rm{M^2},\infty)} \geq 30\%,\\
 \rm{CVG} &=\frac{\Delta^{\rm{DimN}} (\rm{M^2},\rm{s_0})}{\Delta (\rm{M^2},\rm{s_0})} \leq 5\%,
 \end{align}
 where DimN $\geq 10$.  The working intervals of the auxiliary parameters $\rm{M^2}$ and $\rm{s_0}$ are determined by these constraints.  Table~\ref{parameter} presents these intervals for the studied states, along with the corresponding CVG and PC values obtained from the analyses.  For the sake of completeness, as an example, Fig.~\ref{Msqfig} shows the dependence of the magnetic dipole moments on the $\rm{M^2}$ for fixed values of $\rm{s_0}$. The figure illustrates a slight variation in the results obtained in these regions, as required. The results may contain uncertainties due to residual dependencies.
\begin{widetext}

\begin{table}[htp]
	\addtolength{\tabcolsep}{10pt}
	\caption{Working regions of the $\rm{M^2}$ and $\rm{s_0}$ for the magnetic dipole and higher multipole moments.}
	\label{parameter}
		\begin{center}
\begin{tabular}{l|ccccc}
	   \hline\hline
   State&  $\rm{s_0}$~\mbox{[GeV$^2$]}&$\rm{M^2}$~\mbox{[GeV$^2$]}&\mbox{CVG} $[\%]$&\mbox{PC} $[\%]$\\
\hline\hline
$c \bar cuuu$&  $ 24.0-26.0$         &  $ 5.0-7.0 $&< 2.0&30-55\\
$c \bar cddd$ &  $ 24.0-26.0$        &  $ 5.0-7.0 $&< 2.0&32-54\\
$c \bar csss$ &  $ 26.0-28.0$        &  $ 5.5-7.5 $&< 2.0&30-58\\
	   \hline\hline
\end{tabular}
\end{center}
\end{table}

\end{widetext}

Since we have the values of all the input parameters, we can numerically calculate the magnetic dipole, electric quadrupole, and magnetic octupole moments of the hidden-charm pentaquark states. The results of the magnetic dipole and higher multipole moments obtained with these three different interpolating currents are given in Table \ref{table}, together with the errors due to by $\rm{M^2}$ and $\rm{s_0}$, as well as all the numerical parameters used.  As we see from Table \ref{table}, different interpolating currents employed to probe pentaquarks with the same quark content produce varying different results for their magnetic dipole and higher multipole moments at all. 
 This can be interpreted to mean that there are more than one hidden-charm pentaquarks with the same quark content and different magnetic dipole and higher multipole moments (see the discussion following Eq.~(\ref{mqo2})). 
 As mentioned earlier, all these interpolating currents have the same quantum numbers and quark contents, so it leads to the almost degenerate masses for these hidden-charm pentaquarks~\cite{Wang:2015wsa}. However, as we have seen, the results obtained for the magnetic dipole and higher moments are quite sensitive to the configurations and the nature of the diquark-diquark-antiquarks that form the states being studied. An idea of the consistency of our estimates can be obtained by comparing the magnetic dipole and higher multipole moment results obtained with different approaches with the results acquired in this work. 
 
 The quark contents of the experimentally discovered pentaquark states are so far either $c \bar c uud$ or $c \bar c uds$. Therefore, the studies have usually been carried out for states with different quark contents.  In this study, the light quark contents correspond to states with identical quark contents. In the case of the $c \bar c sss$ pentaquark state, there is a study where the result of the magnetic dipole moment obtained by using the quark model in \cite{Wang:2022nqs} and considering the molecular state $\bar D_s^* \Omega_c$ is $-2.19~\mu_N$. As can be seen, this is not consistent with the results of our analysis. For the pentaquark states $c \bar c uuu$ and $c\bar c ddd$ we cannot make a comparison because we have no numerical results to compare.  However, it can be useful to present the existing theoretical results on the molecular configuration of the experimentally discovered $c \bar c uud$ pentaquark states with possible quantum numbers $J^P =3/2^-$ so that the reader will have a better understanding of the results obtained.  In Ref. \cite{Wang:2016dzu}, the magnetic dipole moments of the $\rm{P_c(4380)}$ and $\rm{P_c(4440)}$  states were extracted within the quark model, and the extracted results are $\mu_{\rm{P_c(4380)}}  = 1.357~ \mu_N $ and $\mu_{\rm{P_c(4440)}}  = 3.246~ \mu_N$. In Refs. \cite{ Ozdem:2018qeh, Ozdem:2021ugy}, the magnetic dipole moments of the $\rm{P_c(4380)}$  and $\rm{P_c(4440)}$ states have been predicted by using the QCD light-cone sum rules by assuming that $\bar D^{*-} \Sigma_c^{++}$ and  $\bar D^{*0} \Sigma_c^+$ molecular configurations, respectively. The extracted values are given as $\mu_{\rm{P_c(4380)}}  = 3.35 \pm 1.35~ \mu_N $ and $\mu_{\rm{P_c(4440)}}  = 3.49^{+1.49}_{-1.31}~ \mu_N$.  In Ref. \cite{Ozdem:2024jty}, the magnetic dipole moments of the $\rm{P_c(4380)}$  and $\rm{P_c(4440)}$ states were revisited with the help of the QCD light-cone sum rules by assuming that $\big(\frac{1}{\sqrt{3}}\mid \bar D^0 \Sigma_c^{+*} \rangle \, - \sqrt{\frac{2}{3}}\mid \bar D^- \Sigma_c^{*++} \rangle \big)$ and  $\big(\frac{1}{\sqrt{3}}\mid \bar D^{*0} \Sigma_c^+ \rangle \, - \sqrt{\frac{2}{3}}\mid \bar D^{*-} \Sigma_c^{++} \rangle \big)$ molecular pictures, respectively. The recalculated values are presented as $\mu_{\rm{P_c(4380)}}  = 1.88^{+0.73}_{-0.64}~ \mu_N $ and $\mu_{\rm{P_c(4440)}}  = 0.73^{+0.26}_{-0.24}~ \mu_N$.  As previously stated, the magnetic dipole moment values of the $c \bar c uud$ pentaquarks are given only to make our results understandable for the reader.

 Our final comment on the results of the magnetic dipole and higher multipole moments is an investigation of the violation of the U-symmetry. The effects of the U-symmetry violation have been taken into account in terms of a non-zero s-quark mass and s-quark condensate. From the results we notice that the U-symmetry violation in the magnetic dipole moments is about $11\%$ for the $J^1_{\mu}(x)$ interpolating current, about $19\%$ for the  $J^2_{\mu}(x)$ interpolating current, and about $53\%$ for the $J^3_{\mu}(x)$ interpolating current. In the case of the electric quadrupole moment, we see that the U-symmetry violation is about $19\%$ for the $J^1_{\mu}(x)$ interpolating current, about $21\%$ for the $J^2_{\mu}(x)$ interpolating current, and about $26\%$ for the $J^3_{\mu}(x)$ interpolating current. In the case of the magnetic octupole moment, we realize that the U-symmetry violation is about $17\%$ for the $J^1_{\mu}(x)$ interpolating current, about $13\%$ for the $J^2_{\mu}(x)$ interpolating current, and about $27\%$ for the $J^3_{\mu}(x)$ interpolating current.

  \begin{widetext}
  
\begin{table}[t]
\addtolength{\tabcolsep}{10pt}
	\caption{Numerical values of the magnetic dipole, electric quadrupole, and magnetic octupole moments of the spin-$\frac{3}{2}$ pentaquark states.}
	\label{table}
		\begin{ruledtabular}
\begin{tabular}{l|ccccc}
Parameter& $c \bar cuuu$&$c \bar cddd$&$c \bar csss$ \\
	   \hline\hline 
	   \\
$\mu_{{J_\mu^1}}[\mu_N]$&  $-0.40^{+0.17}_{-0.13}$&$ -0.38^{+0.16}_{-0.13}$&$ -0.34^{+0.12}_{-0.11}$\\
\\
$\mu_{{J_\mu^2}}[\mu_N]$&  $3.51^{+1.52}_{-1.14}$&$ 3.71^{+1.50}_{-1.13}$&$ 3.12^{+1.00}_{-0.79}$\\
\\
$\mu_{{J_\mu^3}}[\mu_N]$&  $1.85^{+0.91}_{-0.68}$&$ 2.30^{+0.84}_{-0.65}$&$1.50^{+0.42}_{-0.34}$\\
\\
	           \hline\hline
	           \\
$\mathcal{Q}_{{J_\mu^1}}(\times 10^{-2})[\mbox{fm}^2]$&  $-0.40^{+0.15}_{-0.12}$&$ -0.37^{+0.15}_{-0.12}$&$ -0.31^{+0.09}_{-0.08}$\\
\\
$\mathcal{Q}_{{J_\mu^2}}(\times 10^{-2})[\mbox{fm}^2]$&  $-2.39^{+0.80}_{-0.62}$&$ 5.96^{+1.80}_{-1.60}$&$ 4.91^{+1.31}_{-1.03}$\\
\\
$\mathcal{Q}_{{J_\mu^3}}(\times 10^{-2})[\mbox{fm}^2]$&  $-9.90^{+3.51}_{-2.48}$&$ 8.23^{+2.79}_{-2.12}$&$ 6.55^{+1.68}_{-1.33}$\\
\\
\hline \hline
\\
 $\mathcal{O}_{{J_\mu^1}}(\times 10^{-3})[\mbox{fm}^3]$&  $-0.022^{+0.008}_{-0.008}$&$-0.020^{+0.007}_{-0.006}$&$ -0.017^{+0.05}_{-0.04}$\\
 \\
$\mathcal{O}_{{J_\mu^2}}(\times 10^{-3})[\mbox{fm}^3]$&  $0.67^{+0.22}_{-0.16}$&$ -1.07^{+0.32}_{-0.27}$&$ -0.95^{+0.18}_{-0.14}$\\
\\
$\mathcal{O}_{{J_\mu^3}}(\times 10^{-3})[\mbox{fm}^3]$&  $2.20^{+0.64}_{-0.49}$&$ -1.77^{+0.48}_{-0.37}$&$ -1.39^{+0.29}_{-0.24}$\\
\end{tabular}
\end{ruledtabular}
\end{table}

\end{widetext}

\section{Summary and concluding remarks}\label{summary}

The theoretical study of nonconventional hadrons has made significant progress with the experimental discovery of more and more nonconventional states involving multiquarks. The magnetic dipole and higher multipole moments of these unconventional states are intrinsic properties that provide valuable insights into their quark constituents. Inspired by this, in the present work, we have explored the magnetic dipole and higher multipole moments of the hidden-charm pentaquarks with quantum number $J^P = 3/2^-$ using the QCD light-cone sum rule method and interpolating currents $J^1_{\mu}(x)$, $J^2_{\mu}(x)$, and $J^3_{\mu}(x)$. As we see from Table \ref{table}, different interpolating currents employed to probe pentaquarks with the same quark content produce varying different results for their magnetic dipole and higher multipole moments at all.  This can be interpreted to mean that there is more than one hidden-charm pentaquark with identical quark content but with different magnetic dipole and higher multipole moments.  An idea of the consistency of our estimates can be obtained by comparing the magnetic dipole and higher multipole moment results obtained with different approaches with the results acquired in this work.
 
 Valuable information about the size and shape of the hadrons can be obtained from the magnetic dipole and higher multipole moments of the pentaquark states with hidden-charm. For the interpretation of the hadron properties in terms of quark-gluon degrees of freedom, the determination of these physical measurables is a crucial step.  Furthermore, the magnetic dipole moment, which can independently probe the pentaquark states with hidden-charm, is an important element in the analysis of the $J/\psi$ photo-production process. If the inner structure of the nonconventional states is elucidated, our understanding of the structure of the subatomic world will be decisively improved, and our insight into the non-perturbative nature of the strong interaction in the low-energy regime will also be decisively improved.

\section{Acknowledgments}
The author acknowledges A. \"{O}zpineci for his valuable contributions to the comments, discussions, and suggestions.

  \begin{widetext}
  %
  \section*{Appendix: Explicit expression for $\rm{F_i}$ form factors}\label{appenda}
 For the $J_{\mu}^1$ interpolating current, the expressions for the $\rm{F_i}$ form factors are given as
%
\begin{align}
 \rm{F_1} &=\frac{e^{\frac{m^2_{P_{\bar cc}}}{\rm{M^2}}}}{\lambda^2_{P_{\bar cc}}}\bigg\{
 \frac {P_ 1 P_ 2^2} {2^{18} 3^6 5 \pi^3}\bigg[90 e_c m_c m_q \rm{I}[0, 1] - 11 e_c \rm{I}[0, 2] - 11 e_q \rm{I}[0, 2]\bigg]\nonumber\\
        &+\frac {m_c P_ 1 P_ 2} {2^{23} 3^7 5 \pi^5} \bigg[ 22 e_c m_ 0^2 (3 m_c + 2 m_q) \rm{I}[0, 2] - 
 64 e_q m_ 0^2 (21 m_c + 11 m_q) \rm{I}[0, 2] + 
 100 e_q (6 m_c + 17 m_q) \rm{I}[0, 3] \nonumber\\
 &- e_c (734 m_c + 827 m_q) \rm{I}[0, 3]\bigg]\nonumber\\
        & + \frac {e_c m_c P_ 2^2} {2^{16} 3^5 5 \pi^3}  \bigg[ e_c (55 m_0^2 m_c m_q \rm{I}[0, 2] - 120 m_c m_q \rm{I}[0, 3] + 9 \rm{I}[0, 4])\bigg]\nonumber
  \end{align}
      \begin{align}
         & + \frac {m_c P_ 1} {2^{27} 3^7 5^2 \pi^7} \bigg[ 120 (57 e_c - 5 e_q) m_c m_q \rm{I}[0, 4] - (187 e_c + 1148 e_q) \rm{I}[0, 5]\bigg]\nonumber\\
             & - \frac {e_c m_c P_ 2} {2^{23} 3^5 5^2 \pi^5}\bigg[e_c (15 m_0^2 (38 m_c - 135 m_q) \rm{I}[0, 4] + 
   32 (-7 m_c + 27 m_q) \rm{I}[0, 5])\bigg]
          -\frac {67 e_c m_c m_q} {2^{24} 3^3 5^3 7  \pi^7} \rm{I}[0, 6]+ \frac{19 e_c \rm{I}[0, 7]}{2^{26} 3^2 5^2 7^2  \pi^7} \bigg\},\\
\rm{F_2} &= m_{P_{\bar cc}}\frac{e^{\frac{m^2_{P_{\bar cc}}}{\rm{M^2}}}}{\lambda^2_{P_{\bar cc}}}\bigg\{
 \frac {P_ 1 P_ 2^2} {2^{18} 3^6 \pi^3}\bigg[ -4 e_q \big(6 m_c m_q \rm{I}[0, 1] - 2 \rm{I}[0, 2] + 7 \rm{I}[1, 1]\big) + 
 e_c \big(72 m_c m_q \rm{I}[0, 1] - 11 \rm{I}[0, 2] 
          + 10 \rm{I}[1, 1]\big)\bigg]\nonumber\\
        &+\frac {m_c P_ 1 P_ 2} {2^{22} 3^7 5^1 \pi^5} \bigg[-5 e_c \bigg ((-166 m_c + 201 m_q) \rm{I}[0, 3] + 
    6 m_ 0^2 ((75 m_c + 8 m_q) \rm{I}[0, 2] - 
       4 (27 m_c + 5 m_q) \rm{I}[1, 1]) \nonumber\\
          & + 
    60 (9 m_c + 2 m_q) \rm{I}[1, 2]\bigg) + 
 2 e_q \bigg (8 (50 m_c + 51 m_q) \rm{I}[0, 3] + 
    15 m_ 0^2 ((33 m_c - 4 m_q) \rm{I}[0, 2] - 
       2 (57 m_c + 28 m_q) \rm{I}[1, 1])  \nonumber\\
          &+ 
    174 (10 m_c + 7 m_q) \rm{I}[1, 2]\bigg)\bigg]\nonumber\\
        & - \frac {e_c m_c P_ 2^2} {2^{15} 3^5 5^1 \pi^3}  \bigg[ 75 m_0^2 m_c m_q \rm{I}[0, 2] - 160 m_c m_q \rm{I}[0, 3] + 9 \rm{I}[0, 4]\bigg]\nonumber\\
         & + \frac {m_c P_ 1} {2^{29} 3^6 5^2 \pi^7} \bigg[ -20 e_q \Big (64 m_c m_q (20 \rm{I}[0, 4] + 49 \rm{I}[1, 3]) + 
    7 (2 \rm{I}[0, 5] - 45 \rm{I}[1, 4])\Big) + 
 e_c \Big (2449 \rm{I}[0, 5]  \nonumber\\
          &+ 1120 m_c m_q (-7 \rm{I}[0, 4] + 36 \rm{I}[1, 3]) + 
    900 \rm{I}[1, 4]\Big)\bigg]\nonumber\\
             & - \frac {e_c m_c P_ 2} {2^{22} 3^4 5^2 \pi^5}\bigg[ 240 m_0^2 (m_c - 3 m_q) \rm{I}[0, 4] + (-92 m_c + 333 m_q) \rm{I}[0, 5]\bigg]
          +\frac {e_c m_c^2 m_q} {2^{21} 3^5 5^2 7 \pi^7} \rm{I}[0, 6]- \frac{223 e_c m_c \rm{I}[0, 7]}{2^{26} 3^2 5^3 7^2 \pi^7} \bigg\},\\
       %
         \rm{F_3} &=4 m_{P_{\bar cc}}\frac{e^{\frac{m^2_{P_{\bar cc}}}{\rm{M^2}}}}{\lambda^2_{P_{\bar cc}}}\bigg\{
         -\frac {e_c m_c P_ 1 P_ 2^2} {2^{16} 3^5 \pi^3}\rm{I}[0, 1]
    \nonumber\\
          &+\frac {m_c P_ 1 P_ 2} {2^{21} 3^6 5 \pi^5} \bigg[4 e_q m_q \Big (-100 m_ 0^2 \rm{I}[0, 1] + 173 \rm{I}[0, 2] + 53 \rm{I}[1, 1]\Big) + 
 e_c \Big (-20 m_ 0^2 (39 m_c - 8 m_q) \rm{I}[0, 1]  \nonumber\\
 &+ 680 m_c \rm{I}[0, 2]- 
    485 m_q \rm{I}[0, 2] - 32 m_q \rm{I}[1, 1]\Big)\bigg]
                     -\frac {e_c m_c^2 m_q P_ 2^2 } {2^{13} 3^4  \pi^3}\bigg[5 m_0^2 \rm{I}[0, 1] - 4 \rm{I}[0, 2]\bigg]\nonumber\\
       &  +\frac {m_c P_ 1 } {2^{28} 3^7 5 7 \pi^7}\bigg[-21 (6400 e_c m_c m_q \rm{I}[0, 3] + 9 (103 e_c + 4 e_q) \rm{I}[0, 4]) - 
 8 (82 e_c + 341 e_q) \rm{I}[1, 3]\bigg]
        \nonumber\\
           &+\frac {e_c m_c P_ 2} {2^{21} 3^4 5 \pi^5} \bigg[-32 m_ 0^2 (4 m_c + 9 m_q) \rm{I}[0, 3] + 9 (4 m_c + 21 m_q) \rm{I}[0, 4]\bigg]
          -\frac {41 e_c m_c^2 m_q } {2^{21} 3^4 5^2 7  \pi^7} \rm{I}[0,5] 
+\frac {191 e_c m_c } {2^{25} 3^3 5^3 7  \pi^7}\rm{I}[0, 6]\bigg\},\\
       \rm{ and}
       \nonumber\\
          \rm{F_4} &=4 m^3_{P_{\bar cc}}\frac{e^{\frac{m^2_{P_{\bar cc}}}{\rm{M^2}}}}{\lambda^2_{P_{\bar cc}}}\bigg\{
           \frac {m_c m_q P_ 1 P_ 2} {2^{18} 3^6 5 \pi^5}(8 e_c - 53 e_q)\rm{I}[0,1]
                 + \frac {m_c P_ 1 } {2^{24} 3^7 5 7 \pi^7}(82 e_c + 341 e_q)\rm{I}[0,3]\bigg\},
        \end{align}
where $P_1 =\langle g_s^2 G^2\rangle$ is gluon condensate, and $P_2 =\langle \bar q q \rangle$ stands for light-quark condensate. It should be noted that the above expressions only include terms that significantly contribute to the numerical values of the magnetic dipole and higher multipole moments. Higher dimensional contributions, although considered in numerical calculations, have not been presented for simplicity.    The function $\rm{I}[n,m]$ is given as 
\begin{align}
 \rm{I}[n,m]&= \int_{4 m_c^2}^{\rm{s_0}} ds~ e^{-s/\rm{M^2}}~
 s^n\,(s-4\,m_c^2)^m.
 \end{align}
 \end{widetext}
\bibliographystyle{apsrev4-1}
\bibliography{PcMM1.bib}
\end{document}